\documentclass[aps,prl,twocolumn,superscriptaddress,longbibliography]{revtex4-1}

\usepackage{graphicx}

\newcommand{\upperRomannumeral}[1]{\uppercase\expandafter{\romannumeral#1}}

\usepackage{hyperref}
\pdfoutput=1
\usepackage{multirow}
\usepackage{array}
\usepackage{amsmath}
\usepackage{bm}
\usepackage{amssymb}
\usepackage{textcomp}
\usepackage{stmaryrd} 
\usepackage{ulem}
\usepackage{gensymb}

\usepackage{import}
\usepackage{color}
\usepackage{verbatim}

\usepackage{float}

\usepackage{sistyle} 

\newcommand\abs[1]{\left| #1 \right|}

\def\XXint#1#2#3{{\setbox0=\hbox{$#1{#2#3}{\int}$}
     \vcenter{\hbox{$#2#3$}}\kern-.5\wd0}}

\begin{document}

\title{Many-body effects in suspended graphene probed through magneto-phonon resonances}

\author{St\'ephane Berciaud}
\email{stephane.berciaud@ipcms.unistra.fr}
\affiliation{Universit\'e de Strasbourg, CNRS, Institut de Physique et Chimie des Mat\'eriaux de Strasbourg, UMR 7504, F-67000 Strasbourg, France}
\affiliation{Institut Universitaire de France, 1 rue Descartes, 75231 Paris cedex 05, France}

\author{Marek Potemski}
\affiliation{Laboratoire National des Champs Magn\'etiques Intenses, CNRS/UJF/UPS/INSA, Grenoble F-38042, France}

\author{Cl\'ement Faugeras}
\email{clement.faugeras@lncmi.cnrs.fr}
\affiliation{Laboratoire National des Champs Magn\'etiques Intenses, CNRS/UJF/UPS/INSA, Grenoble F-38042, France}


\begin{abstract}
We make use of micro-magneto Raman scattering spectroscopy to probe magneto-phonon resonances (MPR) in suspended mono- to penta-layer graphene. MPR correspond to avoided crossings between zone-center optical phonons (G-mode) and optically-active inter Landau level (LL) transitions and provide a tool to perform LL spectroscopy at a fixed energy ($\approx 197~\rm{meV}$) set by the G-mode phonon. Using a single-particle effective bilayer model, we readily extract the velocity parameter associated with each MPR. A single velocity parameter slightly above the bulk graphite value suffices to fit all MPR for $N\geq2$ layer systems. In contrast, in monolayer graphene, we find that the velocity parameter increases significantly from $(1.23\pm 0.01) \times 10^6~\mathrm{m.s^{-1}}$ up to $(1.45\pm0.02) \times 10^6~\mathrm{m.s^{-1}}$ as the first to third optically-active inter LL transition couple to the G-mode phonon. This result is understood as a signature of enhanced many-body effects in unscreened graphene.
\end{abstract}


\maketitle

Pristine suspended monolayer graphene is a well-defined, unscreened two-dimensional electronic system, in which a wealth of intriguing electronic~\cite{Bolotin2008,Du2008,Bolotin2009,Elias2011,Faugeras2015}, optical~\cite{Mak2012} and mechanical~\cite{Castellanos-Gomez2015} properties have been uncovered. In particular, spectacular deviations from a simple one-electron picture of graphene's band structure (i.e., the Dirac cones) emerge at low carrier densities (below a few $10^{11}~\mathrm{cm^{-2}}$). Due to electron-electron interactions, the velocity parameter diverges logarithmically as the Fermi energy approaches the Dirac point~\cite{Elias2011}, reaching values,  well above the Fermi velocities of bulk graphite and supported graphene ($\sim 1\times 10^6 \mathrm {\; m.s^{-1}}$)~\cite{Basov2014}. 


In the presence of a transverse magnetic field $B$, the electronic states of graphene merge into discrete, highly degenerate Landau Levels (LL)~\cite{Orlita2010}. The energy and lifetime of LL can be probed using magneto-transport measurements~\cite{Bolotin2008,Elias2011}, scanning tunnelling spectroscopy~\cite{Li2009} and magneto-optical spectroscopies~\cite{Orlita2010,Chen2014}. Recently, micro-magneto-Raman spectroscopy (MMRS)~\cite{Faugeras2009,Faugeras2011,Berciaud2014,Kim2013,Neumann2015a,Kazimierczuk2019} has been employed to probe the electronic dispersion of suspended graphene layers~\cite{Berciaud2014} and also to demonstrate that electronic excitations between LL  may be strongly affected by many-body effects.  First, as $B$ decreases, the energy of a given LL also decreases and electron-electron interactions lead to a logarithmic divergence of the corresponding velocity parameter (hereafter denoted $v_{F}$ for simplicity)~\cite{Faugeras2015,Neumann2015a,Sonntag2018}. Second, inter LL excitations lead to the formation of magneto-excitons, whose binding energies depend  on the index $n$ of the electron and hole LL they arise from, leading to $n$-dependent $v_{\rm F}$~\cite{Faugeras2015}. 

Thus far, Raman signatures of many-body effects have predominantly appeared on the electronic Raman scattering response of graphene~\cite{Riccardi2016} under a transverse magnetic field~\cite{Berciaud2014,Faugeras2015}. Recently, many-body effects have also been unveiled by monitoring magneto-phonon resonances~\cite{Faugeras2009} (MPR) between optically-active inter-LL transitions~\cite{Russel2018} and zone-center optical phonons (i.e., Raman G-mode phonons~\cite{Ferrari2013}) in graphene encapsulated in hexagonal boron nitride (BN) films~\cite{Neumann2015a}. As recently suggested by Sonntag \textit{et al.}~\cite{Sonntag2018}, more prominent effects are expected in suspended graphene, where electron-electron interactions are minimally screened.

\begin{figure*}[!ht]
\begin{center}
\includegraphics[width=0.6\linewidth]{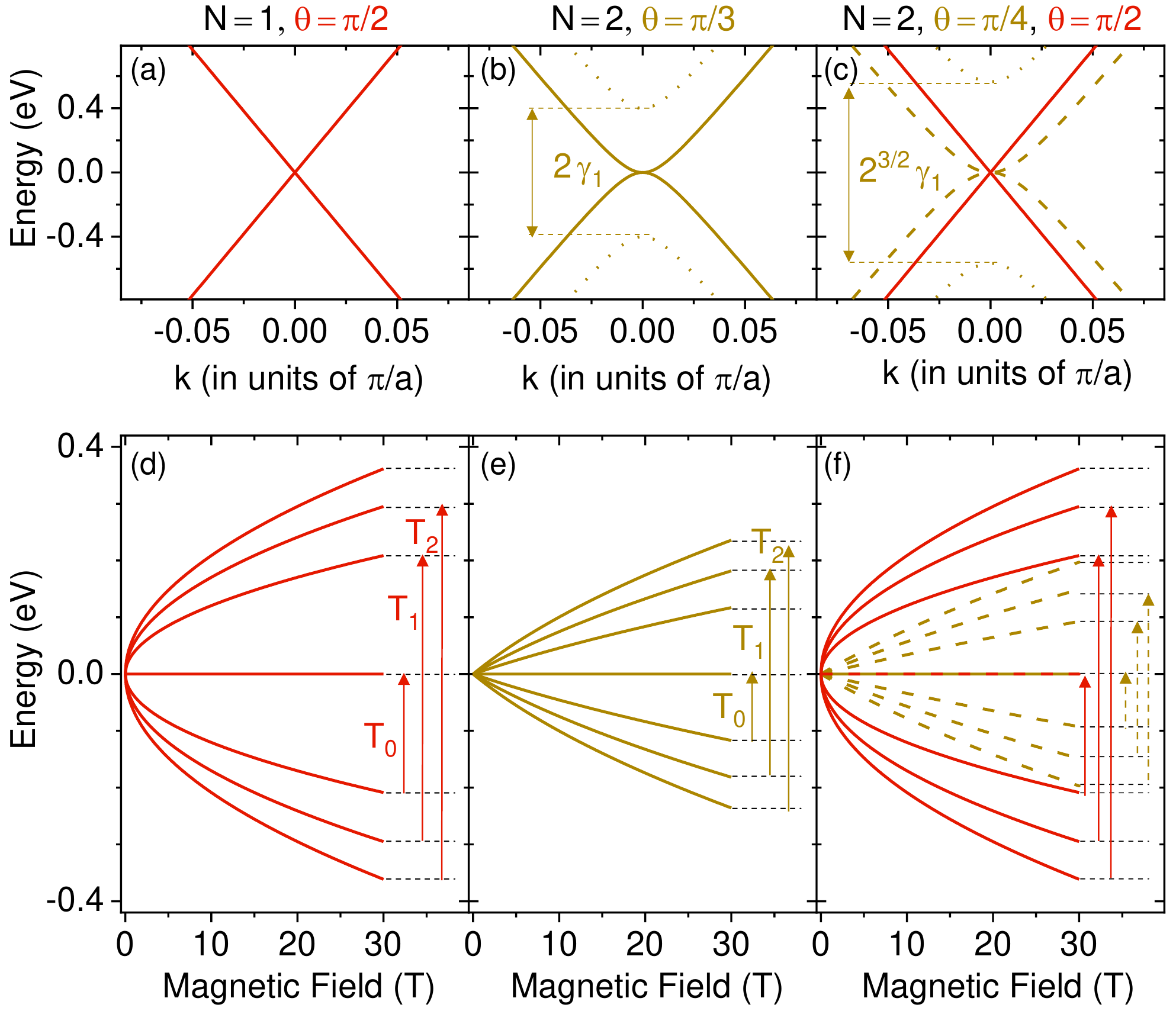}
\caption{Single-particle low-energy electronic structure and Landau levels of mono-, bi- and trilayer graphene. The electronic dispersions obtained from the effective bilayer model described in the text are shown in (a-c), respectively. The corresponding dispersion of the Landau levels arising from the gapless bands are shown in (d,e,f) for mono-, bi- and trilayer graphene, respectively. Red  lines in (d) and (f) correspond to monolayer-like Landau levels $\left(\theta=\pi/2\right)$, while dark yellow lines in (e,f) correspond to Landau levels arising from effective bilayers obtained at quantized values of $\theta\neq  \pi/2$. The vertical arrows indicate the optically allowed inter LL transitions $T_n$ that give rise to the magneto-phonon resonances. The calculations are performed with $v_{\mathrm F}=1.05 \times 10^6~\mathrm m/s$ and $\gamma_1=400~\mathrm meV$.  }
\label{Fig1}
\end{center}
\end{figure*}

In this letter, we report the results of MMRS measurements performed on suspended mono- to pentalayer graphene. For each number of layers $N$, we resolve a set of well-defined MPR. Using a single-particle effective bilayer model, we readily extract $v_{\rm F}$ associated with each  MPR. While a single parameter slightly above the bulk graphite value ($\gtrsim 1.05~\times 10^6~\mathrm{m.s^{-1}}$) suffices to fit all MPR for $N\geq2$ layer systems, We find that $v_{\rm F}$ increases significantly up to $\approx 1.45 \times 10^6~\mathrm{m.s^{-1}}$ in monolayer graphene. This result is understood as a signature of enhanced many-body effects in unscreened graphene.


The low-energy electronic electronic bands and magneto-optical response of mono- and $N-$layer graphene have been extensively discussed using an effective bilayer model~\cite{Latil2006,Koshino2007,Partoens2007,Orlita2009a,Mak2010,Berciaud2014}. To lowest order, this model uses only two parameters, namely the single-particle Fermi velocity $v_{\mathrm{F}}=3/2 a \gamma_0/\hbar$, where $\gamma_0$ is the nearest neighbor hopping parameter, $a=0.142~\mathrm{nm}$ is the \textsc{c--c} bond length, and the nearest neighbor interlayer coupling constant $\gamma_1$. The band structure of Bernal- or equivalently AB-stacked $N\geq2$ layer systems exhibits $2N$ bands, including one pair of linear (monolayer-like) bands only present for odd $N$ and a set of $\left\lfloor N/2\right\rfloor$ effective bilayer bands (here, $\left\lfloor \right\rfloor$ denotes the integer part) with a rescaled  $\tilde{\gamma}_1=2\gamma_1\!\cos\theta$~\cite{Koshino2007,Partoens2007,Orlita2009a,Mak2010}, which corresponds to half the energy gap between their split-off bands. The angle $\theta$ represent the quantized transverse momenta $\theta=k_z~c/2$ at which two-dimensional cuts are made in the three-dimensional electronic dispersion of graphite, with $c/2=0.34~\mathrm{nm}$ the interlayer separation in  bulk graphite. The band structure of mono- to trilayer graphene and corresponding values of $\tilde{\gamma}_1$ and $\theta$ are shown in Fig.~\ref{Fig1}a-c.

\begin{figure*}[!th]
\begin{center}
\includegraphics[width=0.66\linewidth]{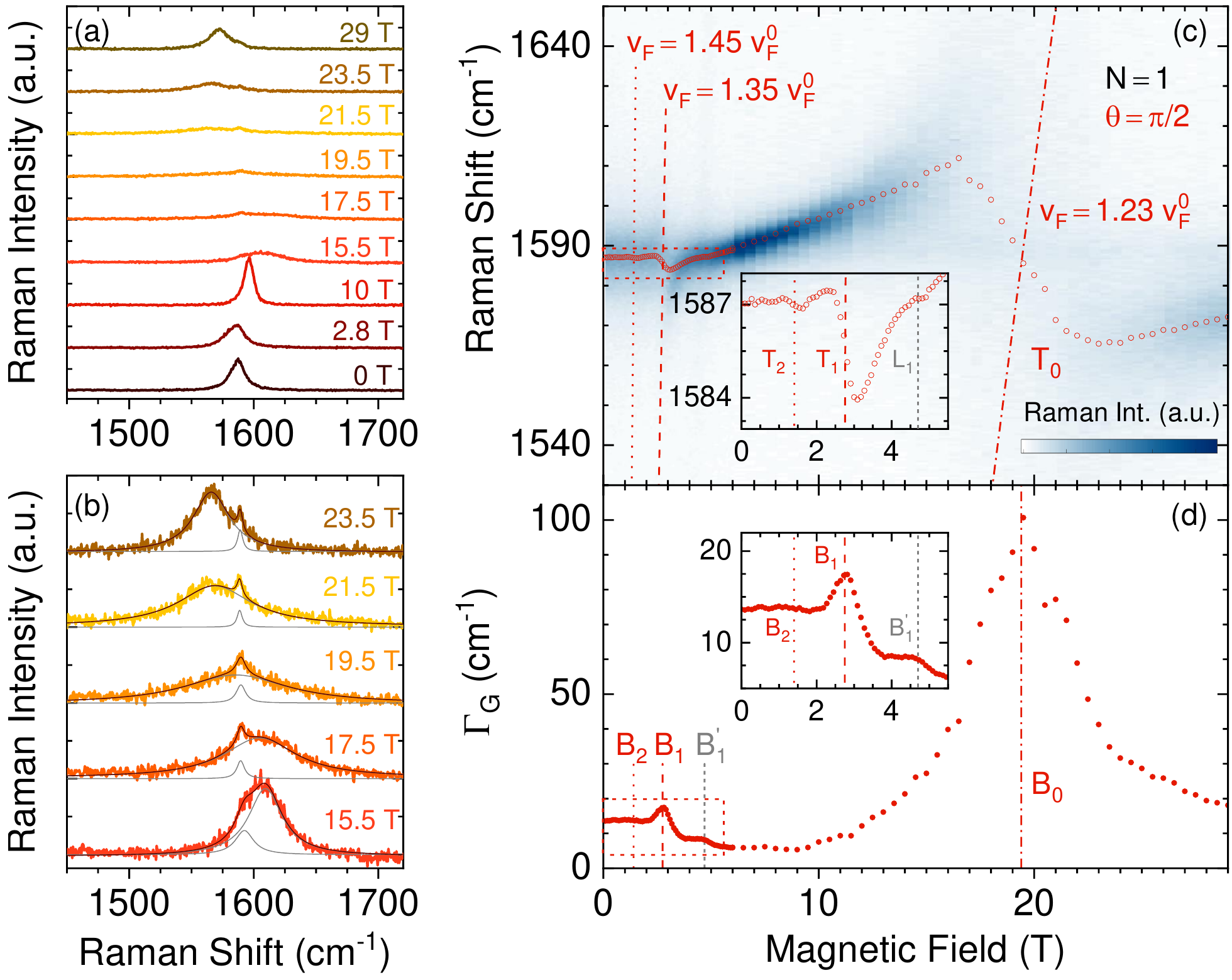}
\caption{Magneto-phonon resonances in suspended monolayer graphene. (a,b) Raman G-mode spectra at selected values of the transverse magnetic field $B$. The spectra are acquired for the same duration and incident laser power and offset for clarity. The dark and light gray lines are Lorentzian fits to the data. (c) G-mode frequency $\omega_{\mathrm G}$ (red circles), extracted from Lorentzian fits of the G-mode spectra, overlaid on a false-color plot of the Raman spectra as a function of $B$.  The red and grey lines are calculated dispersions of the $T_n$ (optically allowed) and of the $L_1$ (Raman allowed) transitions, respectively. The velocity parameters (estimated at $\hbar\omega_{\rm G}$) associated with the three observed MPR are indicated in units of $v^0_{\mathrm F}=1.00 \times 10^{6}~\mathrm{m.s^{-1}}$.  (d) Lorentzian full-width at half maximum $\Gamma_{\mathrm G}$ as a function of the magnetic field. The resonant fields $B_n$ are indicated. The insets in (c) and (d) show  close-ups on the regions boxed with thin dotted lines.}
\label{Fig2}
\end{center}
\end{figure*}

\begin{figure*}[!ht]
\begin{center}
\includegraphics[width=0.7\linewidth]{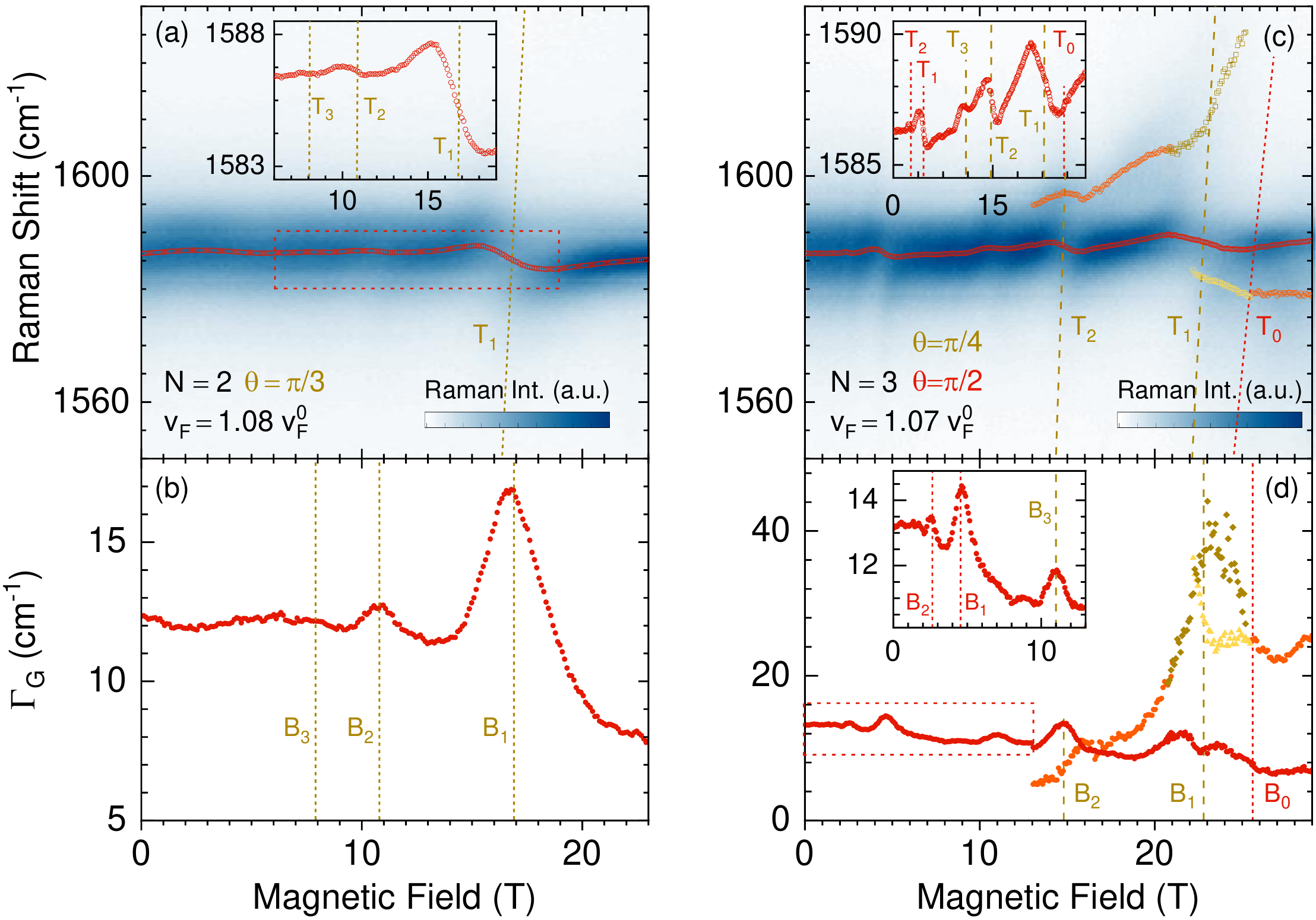}
\caption{Magneto-phonon resonances in suspended bi- (a,b) and trilayer (c,d) graphene. (a,c) G-mode frequency $\omega_{\mathrm G}$ (red circles), extracted from a Lorentzian fit of the G-mode spectra, overlaid on a false-color plot of the Raman spectra as a function of the transverse magnetic field $B$.  The red and dark yellow dashed lines are calculated dispersions of the $T_n$ transitions for monolayer-like and effective bilayer bands. The single $v_{\mathrm F}$ extracted from all MPR is indicated in both cases in units of $v^0_{\mathrm F}=1.00~\times 10^{6}~\mathrm{m.s^{-1}}$. (b,d) Lorentzian full-width at half maximum $\Gamma_{\mathrm G}$ as a function of the magnetic field. The resonant fields $B_n$ are indicated with color-coded values of $\theta$. In the trilayer case (c, d), the G-mode spectra are fit to a double Lorentzian (red and orange symbols) for $13~\mathrm{T} < B < 21~\mathrm {T}$ and $26~\mathrm{T} <B<29 ~\mathrm{T}$ and to a triple Lorentzian (red, yellow and dark yellow) for $21~\mathrm{T} <B<26~\mathrm{T} $ near the  first monolayer-like and second bilayer-like MPR. The insets in (a,b,d) show close-ups on the regions boxed with red dotted lines in (a,d) and on the most prominent G-mode sub-feature in (c).}
\label{Fig3}
\end{center}
\end{figure*}

In the presence of a transverse magnetic field, the effective mono- and bilayers give rise to independent sets of Landau fans. The energy  $\mathcal{L}^{\theta}_{n}$ $\left(\mathcal{L}^{\theta}_{-n}=-\mathcal{L}^{\theta}_{n}\right)$ of the $n^{th}$  electron ($n\geq0$) or hole ($n\leq0$) LL arising from the gapless bands in an effective bilayer writes~\cite{McCann2006a,Orlita2010}:

\begin{widetext}
\begin{equation}
\mathcal{L}^{\theta}_{\left|n\right|}=\sqrt{\frac{\tilde{\gamma}_1^2}{2}+\left(\left|n\right|+\frac{1}{2}\right)E_1^2-\sqrt{\frac{\tilde{\gamma}_1^4}{4}+\left(\left|n\right|+\frac{1}{2}\right)E_1^2\tilde{\gamma}_1^2+\frac{E_1^4}{4}}},
\label{eqLL1}
\end{equation}
\end{widetext}
where $E_1=v_\mathrm{F}\sqrt{2e\hbar B}$.  At $\theta=\pi/2$, one obtains the well-known LL fan of monolayer graphene $\mathcal{L}^{\pi/2}_{\left|n\right|}=\sqrt{n}E_1$. For $\theta\neq\pi/2$, nearly linear scalings of $\mathcal{L}^{\theta}_{n}(B)$, characteristic of effective bilayers are observed. The corresponding LL fans are shown in~Fig.~\ref{Fig1}d-f for mono to trilayer graphene, respectively. As previously established \cite{Kashuba2009,Mucha-Kruczynski2010,Kossacki2011}, optically-allowed transitions, such that $\delta \left| n \right| = \pm 1$ couple to the graphene G-mode phonons (with energy $\hbar\omega_{\rm G}\approx 1585~\mathrm {cm}^{-1}$ or equivalently $\approx 197~\mathrm{meV}$) and give rise to series of MPR~\cite{Ando2007a,Goerbig2007,Faugeras2009}, as indicated by the vertical arrows in Fig.~\ref{Fig1}d-f. In contrast, symmetric inter LL transitions ($\delta \left| n \right| = 0$, denoted $L_n$, with $n\geq 1$) are Raman-allowed~\cite{Faugeras2011,Berciaud2014,Faugeras2015} but are not expected to couple to optical phonons~\cite{Kashuba2009}. In the following, the energy of the optically-active inter LL transitions associated with a given value of $\theta$ will be simply denoted

\begin{equation}
T_n=\abs{\mathcal{L}_{\pm n+1}^{\theta}-\mathcal{L}_{\mp (n)}^{\theta}},~ n\geq0
\label{eqLL}
\end{equation}
and the $(n+1)$\textsuperscript{th} MPR ($n\geq0$) occurs when $T_{n}=\hbar\omega_{\rm G}$.

\begin{figure*}[!ht]
\begin{center}
\includegraphics[width=0.7\linewidth]{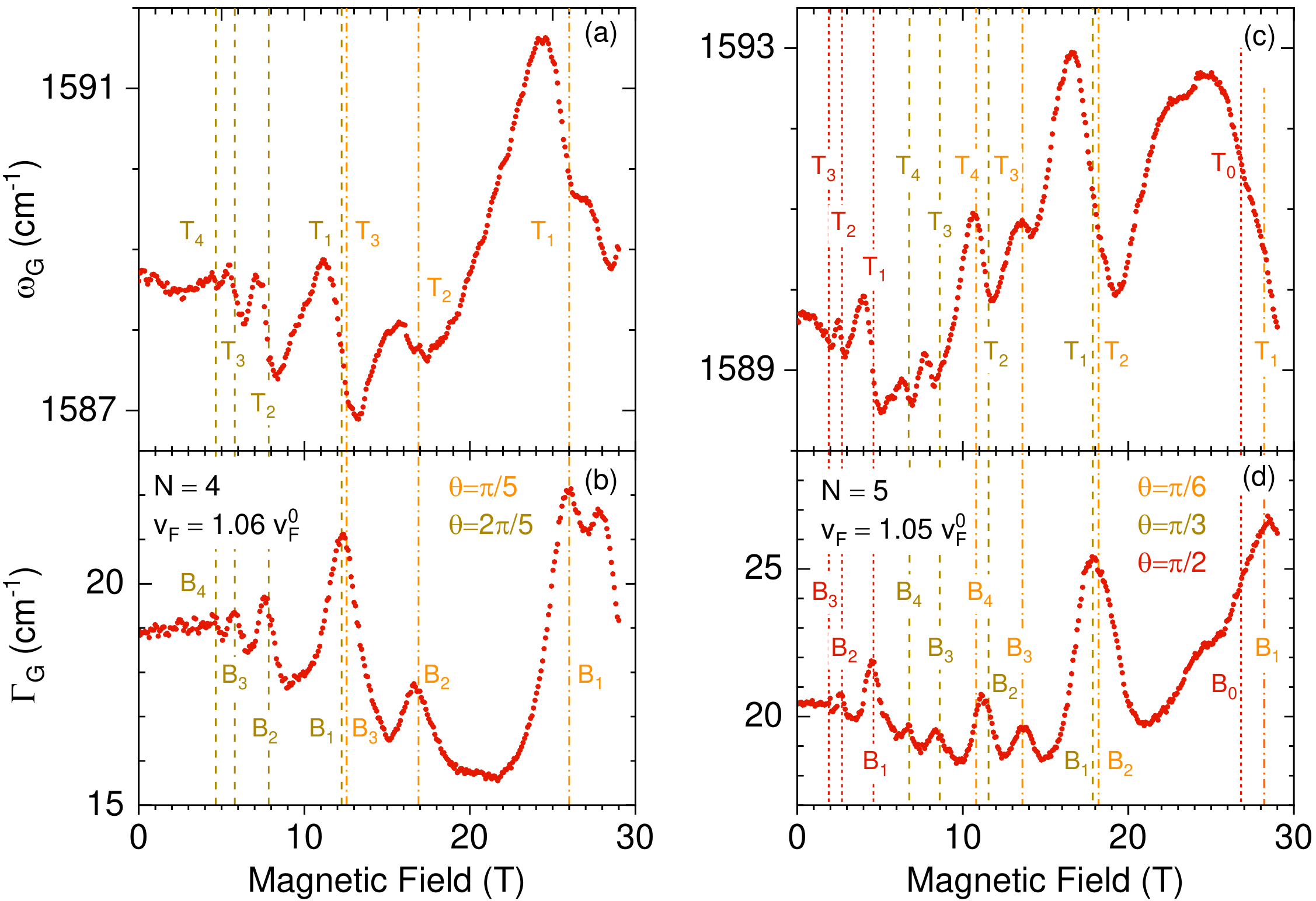}
\caption{Magneto-phonon resonances in suspended tetra- (a,b) and pentalayer (c,d) graphene. (a,c) G-mode frequency $\omega_{\mathrm G}$ (red circles), extracted from a Lorentzian fit of the G-mode spectra as a function of the transverse magnetic field $B$.  The red, dark yellow and orange dashed lines are calculated dispersions of the $T_n$ transitions for monolayer-like and effective bilayer bands with color-coded values of $\theta$. The single $v_{\mathrm F}$ extracted from all MPR is indicated in both cases in units of $v^0_{\mathrm F}=1.00\times 10^{6}~\mathrm{m.s^{-1}}$. (b,d) Lorentzian full-width at half maximum $\Gamma_{\mathrm G}$ as a function of the magnetic field. The resonant fields $B_n$ are indicated with color-coded values of $\theta$.}
\label{Fig4}
\end{center}
\end{figure*}

The suspended graphene samples investigated here were prepared by mechanical cleavage of Bernal stacked natural graphite and dry transfer onto  $8~\mathrm{\mu m}$-diameter pits etched in Si/SiO$_2$ substrates as described in details elsewhere~\cite{Berciaud2009,Metten2014}. Such pristine, minimally screened graphene layers have been shown to exhibit negligible residual doping and charge inhomogeneity, making them ideally suited for MMRS studies~\cite{Berciaud2014,Faugeras2015}. The samples were held in a home-built magneto-optical cryostat where MMRS measurements were performed in a back-scattering geometry under transverse magnetic fields up to 29~T.  A cw laser beam at 514~nm was focused onto a $\sim 2~\mathrm{\mu m}$-diameter spot. A cross-circular polarization configuration was implemented, where incoming and Raman scattered photons have opposite helicity. In these conditions, the magneto-Raman spectra are dominated by MPR~\cite{Kuhne2012,Kashuba2009}. We have employed laser powers $\sim 1~\mathrm{\mu W}$ at our samples to minimise laser-induced heating effects while still having a sufficiently large MMRS signal. All measurements were performed at a base temperature of 4~K. Raman G-mode spectra were fit to Lorentzian functions and the peak frequencies and full-width at half maximum are determined with an experimental error that is typically smaller than the symbol size.

Figure \ref{Fig2} shows MMRS data for a suspended graphene monolayer with minimal native doping (spatially averaged over our laser spot size) well below $10^{12}~\mathrm {cm^{-2}}$. MPR appear as anti-crossings involving significant frequency shifts and line broadening of the G-mode feature. Due to electronic broadening in graphene~\cite{Bostwick2007,Neumann2016}, anti-crossings are not fully resolved, particularly so at low field. The third and second MPR, involving transitions $T_2$ and $T_1$ appear at $B_2=1.45\pm 0.1~\rm T$ and $B_1=2.75\pm 0.1~\rm T$, respectively. Remarkably, for the first time in suspended graphene, we report the $n=0$ MPR at $B_0=19.4\pm0.1~\rm T$. The prominence of the $T_0$ MPR originates from the fact that the coupling strength in MPR theory~\cite{Goerbig2007} grows with the LL degeneracy, which is proportional to $B$. As a result, MPR involving LL with higher $n$ are fainter since these resonances occur at lower $B$.

The resonant fields observed here are considerably lower than in substrate-supported systems~\cite{Faugeras2009,Kim2013,Neumann2015a} as discussed in more details later. As shown in Fig.~\ref{Fig2}b, near $B_0$, the G-mode exhibits a double-Lorentzian lineshape, with a broad, prominent feature that shows an avoided crossing with the $T_0$ transition and a fainter sub-feature (not displayed in Fig.~\ref{Fig2}c,d) that is virtually independent on $B$.  The integrated intensity of this \textit{uncoupled} sub-feature is about one order of magnitude smaller than that of the coupled G-mode feature (Fig.~\ref{Fig2}b); its frequency at $1589.2 \pm 0.8~\rm{cm}^{-1}$ is slightly upshifted and its full-width at half maximum $(6.5\pm 1.5 \mathrm{cm^{-1}})$ is significantly lower as compared with the G-mode feature recorded at $B=0~\rm T$ ($\omega_{\rm G}=1587.0 \pm 0.1 ~\mathrm{cm}^{-1},~ \Gamma_{\rm G}=13.6 \pm 0.1 ~\mathrm{cm}^{-1}$). These characteristics suggest a slight doping with a Fermi level slightly above $\hbar \omega_{\rm G}/2$ (i.e., $\sim 100~\rm{meV}$)~\cite{Pisana2007,Yan2007,Froehlicher2015}. We therefore suggest that local inhomogenities at the nanometer scale may result in an uncoupled G-mode sub-feature due to Pauli blocking of the $T_0$ transition~\cite{Kim2013,Leszczynski2014}. Importantly, Ref.~\onlinecite{Kim2013} reported a similar uncoupled G-mode feature in SiO$_2$-supported graphene, however with a large integrated intensity, comparable with that of the coupled G-mode feature that follows an anti-crossing with the $T_0$ transition. Here, the much lower spectral weight of the uncoupled G-mode sub-feature illustrates the superior spatial homogeneity of suspended graphene.~We suggest that MMRS near the $n=0$ MPR can be used as a quantitative probe of residual electron and hole puddles in high-quality graphene. 

In Fig.~\ref{Fig2}c, one can clearly see an increase of the phonon linewidth, a signature of a variation of the phonon lifetime induced by an interaction, at $B_1^{\prime}=4.7\pm0.1~\rm T$. This resonance implies the G-mode phonon and the symmetric inter LL excitation $L_1$. Although symmetric electronic excitations should not couple to the optical phonons~\cite{Kashuba2009}, a similar MPR has been observed in graphene on graphite~\cite{Faugeras2011} and on graphene encapsulated in hexagonal Boron Nitride (BN)~\cite{Neumann2015a}. Interestingly, both in suspended graphene and in BN-encapsulated graphene, these nominally forbidden MPR are much fainter than the allowed MPR involving $T_1$ and have similar strength. Here, since we investigate suspended graphene, substrate-induced effects cannot account for this unexpected coupling. Our results point towards an intrinsic effect, possibly originating from LL mixing induced by Coulomb interactions~\cite{Roldan2010}.


We now consider how MPR evolve for thicker graphene stacks, as the electronic structure gets increasingly complex. Figure~\ref{Fig3} shows MPR data for suspended bilayer and trilayer graphene. Data for tetra- and penta-layer graphene are reported in Fig.~\ref{Fig4}. These data were acquired using a lower spectral resolution that for $N=1,2,3$. As a result, we have not attempted to resolve signatures of pronounced anti-crossings at high fields for $N=4$ and $N=5$.
For $N=2$, three MPR are resolved at $B_3=7.9\pm0.1~\mathrm T$, $B_2=10.8\pm0.1~\mathrm T$, $B_1=16.6\pm0.1~\mathrm T$ and are assigned to the $T_3$, $T_2$ and $T_1$ MPR of bilayer graphene ($\theta=\pi/3$), respectively. In trilayer graphene (Fig.~\ref{Fig3}c,d), we are able to resolve the first three monolayer-like MPR (involving transitions $T_0$, $T_1$, $T_2$ for $\theta=\pi/2$) and the  MPR associated with $T_1$, $T_2$, $T_3$ for the effective bilayer ($\theta=\pi/4$). At high fields $B>14~\mathrm T$, the G-mode lineshape becomes more complex due to the overlap between the second bilayer-like MPR ($T_1$, $\theta=\pi/4$) and first monolayer-like MPR ($T_0$, $\theta=\pi/2$). These spectra are well fit to a bilayer Lorentzian form for for $13 ~\mathrm T<B<21~ \mathrm T$ and $26 ~\mathrm T<B<29~ \mathrm T$ and to a triple Lorentzian (red, yellow and dark yellow) for $21 ~\mathrm T<B<26~ \mathrm T$. The narrow, central feature (red symbols in Fig.~\ref{Fig3}c) is assigned to the second bilayer-like MPR ($T_1$, $\theta=\pi/4$), while the broader and shifted features (orange symbols for double-Lorentzian fits then dark yellow and yellow symbols for triple-Lorentzian fits in Fig.~\ref{Fig3}c)  are assigned to the first monolayer-like MPR ($T_0$, $\theta=\pi/2$).

In tetralayer graphene (Fig.~\ref{Fig4}a,b), under transverse magnetic fields up to $B=29~\mathrm T$, we observe six MPR associated with gapless effective bilayer bands at  $\theta=\pi/5$ and $\theta=2\pi/5$, respectively~\cite{Faugeras2012}. Up to nine MPR are observed in the same window for pentalayer graphene (Fig.~\ref{Fig4}c,d). These MPR are associated with the effective monolayer ($\theta=\pi/2$), with two other subsets of MPR stemming from the gapless effective bilayer bands at $\theta=\pi/3$ and $\theta=\pi/6$, respectively. Let us note that as the number of electronic subbands increase, MPR involving distinct values of $\theta$ and $n$ may be nearly degenerate.  As a result, we are not able to clearly resolve $T_1,~\theta=2\pi/5$ and $T_3,~\theta=\pi/5$ near 12~T for $N=4$ (Fig.~\ref{Fig4}c,d). Similarly, for  $N=5$ (Fig.~\ref{Fig4}c,d), the MPR associated with  $T_2,~\theta=\pi/3$ and $T_4,~\theta=\pi/6$ near 11~T as well as with $T_0,~\theta=\pi/2$ and $T_1,~\theta=\pi/6$ near 27-28~T appear as single, slightly broadened MPR.  Thus, the growing complexity of the LL spectrum with increasing $N$ prevents detailed studies of MPR for $N>5$, until the limit of the LL fan diagram of bulk graphite is reached.


\begin{figure}[!th]
\begin{center}
\includegraphics[width=\linewidth]{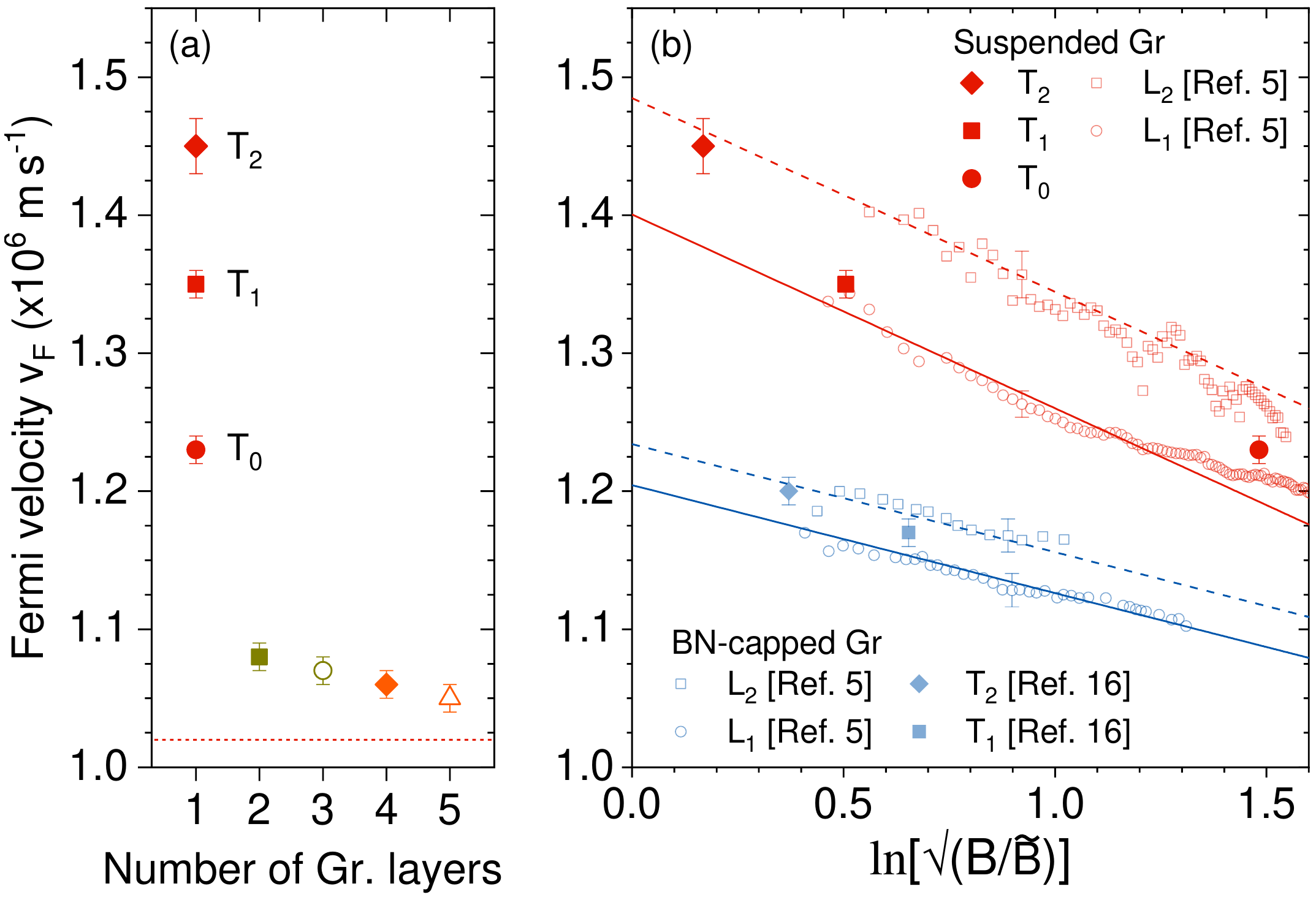}
\caption{(a) Velocity parameters $v_{\rm F}$ extracted from the magneto-phonon resonances in Fig.~\ref{Fig2}-\ref{Fig4} as a function of $N$, the number of graphene layers.  (b) Velocity parameters in suspended monolayer graphene  (filled red symbols) as a function of $\ln{\sqrt{B/\widetilde{B}}}$, with $\widetilde{B}=1~\rm T$. These velocity parameters are compared with those obtained from electronic Raman scattering measurements reproduced from Ref.~\onlinecite{Faugeras2015}), with the symmetric $L_{1}$ and $L_{2}$ transitions in the same sample (open red symbols) and in a graphene sample encapsulated in hexagonal boron nitride (BN, open blue symbols). The solid and dashed lines are theoretical calculations reproduced from Ref.~\onlinecite{Faugeras2015}. These calculations include electron-electron interactions (i.e., the electron self-energy) and magneto-excitonic effects. For comparison, we show (blue symbols) the values of $v_{\rm F}$ measured in Ref.~\onlinecite{Neumann2015a} in BN-capped graphene.}
\label{Fig5}
\end{center}
\end{figure}


As shown in Fig.~\ref{Fig5}a, from the data in Fig.~\ref{Fig2}-\ref{Fig4} and using the effective bilayer model (Eq.~\eqref{eqLL}), we can now determine, for each MPR, the velocity parameters $v_{\rm{F},n}$ evaluated at resonant fields $B_n$. For simplicity, we have considered a constant $\gamma_1=400~\rm {meV}$ for all $N$ and have attempted a global fit of all the $B_n$ measured for a given $N$. For $N=2$ to $N=5$, This procedure yields a very good agreement between  Eq.~\eqref{eqLL} and our experiments with a constant $v_{\mathrm F}$ that decreases smoothly from $(1.08\pm0.01)\times 10^{6}\;\mathrm{m.s^{-1}}$ for $N=2$ down to $(1.05\pm0.01)\times 10^{6}\;\mathrm{m.s^{-1}}$ for $N=5$. These values converge towards the bulk graphite value of $(1.02\pm0.01)\times 10^{6}\;\mathrm{m.s^{-1}}$~\cite{Orlita2009a} and are in excellent agreement with previous electronic Raman measurements on the same samples~\cite{Berciaud2014}.

In contrast, a single velocity parameter would grossly fail to fit the three observed MPR (involving $T_n$, with $n=0,1,2$) in monolayer graphene. Instead, using Eq.~\eqref{eqLL}, we determine that $v_{\mathrm F}$ increases significantly from $v_{\rm F,0}=(1.23\pm0.01)\times 10^6~\mathrm{m.s^{-1}}$ up to $v_{\rm F,2}= (1.45\pm0.02)\times 10^6~\mathrm{m.s^{-1}}$ as $B_n$ decreases from 19.4~T down to  $1.40 ~\mathrm T$. Interestingly, our value of $v_{\rm F,1}=(1.35\pm 0.1)~\times 10^6\, \mathrm{m.s^{-1}}$ inferred from the $n=1$ MPR exactly matches a recent report in neutral suspended graphene grown by chemical vapor deposition and carefully cleaned after transfer (Ref.~\onlinecite{Sonntag2018}).

The rise of $v_{\mathrm{F},n}$ as $B_n$ decreases is qualitatively consistent with our previous measurements~\cite{Faugeras2015} and may arise from the combination of two effects: i) a logarithmic divergence of $v_\mathrm{F}$ as $B_n$ decreases  (self-energy) and ii) magneto-Coulomb binding, leading $n$-dependent $v_{\mathrm{F}}$ associated with $T_n$ or $L_n$ transitions. We should stress that MPR probe charge carriers at a fixed energy set by $\hbar \omega_{\rm G}=T_n$. However, in a one-electron picture, MPR occur at  $B_n \propto \left(\sqrt{n+1}+\sqrt{n}\right)^{-1}$ for $N=1$  (Eq.~\eqref{eqLL}). Previous investigations of $L_n$ transitions, in suspended graphene (Ref.~\onlinecite{Faugeras2015}) and of $T_n$ transitions in BN-encapsulated graphene (Ref.~\onlinecite{Russel2018}) indicate that at a given $B$, $v_{\mathrm{F}}$ increases with $n$, up to $n=3$ (see also Fig.~\ref{Fig5}b). Our data are in line with these results, since $v_{\mathrm{F},0}<v_{\mathrm{F},1}<v_{\mathrm{F},2}$ (Fig.~\ref{Fig2}c). However, Ref.~\onlinecite{Russel2018} also reports a slight decrease of $v_{\mathrm{F}}$ at fixed $B$ for higher order $T_n$ (with $n=3,4,5$). In the wake of recent studies of filling-factor dependent many-body effects in graphene~\cite{Sokolik2018,Shizuya2018,Russel2018,Sonntag2018}, further efforts are needed to separate the contributions of self-energy and magneto-Coulomb binding to the renormalization of $v_{\mathrm{F}}$.

Remarkably, in keeping with our previous measurements of $L_n$ transitions~\cite{Berciaud2014}, many-body effects are essentially observed in monolayer graphene and nearly vanish for $N\geq2$. Such observations are consistent with recent works showing that the reduction of the exciton binding energy in an atomically thin semiconductor coupled to graphene layers is readily maximized with one graphene monolayer and does not increase further if a bilayer graphene is used instead~\cite{Lorchat2020}. In addition, the parabolic dispersion of massive fermions in effective bilayers (Fig.~\ref{Fig1}b,c) creates a finite density of states at the Dirac point, which may be sufficient to quench electron-electron interactions and the subsequent renormalization of electronic bands. Here also, additional investigations of many-body effects are required to account for the abrupt transition between the mono and bilayer cases~\cite{Shizuya2010,Shizuya2011}.

In conclusion, we have demonstrated that magneto-phonon resonances provide invaluable fingerprints of the low-energy electronic dispersion of suspended graphene layers, which are in excellent quantitative agreement with a one-electron effective bilayer model for $N\geq2$. In the monolayer limit, electrons undergo minimal screening, leading to pronounced many-body effects which, to experimental accuracy, do not yield sizeable experimental signatures for $N\geq2$. Many-body effects in graphene and Dirac fermions in general continue to attract considerable interest and future directions include the interplay between of  many-body effects and electronic lifetime in graphene and related systems~\cite{Tang2018}.



~

{\bf Acknowledgements}

We are grateful to D.M. Basko for fruitful discussions. We thank I. Breslavetz, the StNano clean room staff, M. Romeo, F. Chevrier, A. Boulard and the IPCMS workshop for technical support. We acknowledge financial support from the Agence Nationale de Recherche (ANR) under grants H2DH ANR-15-CE24-0016, ANR-17-CE24-0030, 2D-POEM ANR-18-ERC1-0009, the Labex NIE project ANR-11-LABX-0058-NIE and the USIAS GOLEM project, within the Investissement d'Avenir program ANR-10-IDEX-0002-02. M.P. and C.F. acknowledge support from the EC Graphene Flagship project (no. 604391). Part of this work was  performed at the LNCMI-Grenoble, a member of the European Magnetic Field Laboratory (EMFL).
~

%




\end{document}